\documentclass[11pt]{amsart}

\usepackage{amssymb,amsmath,amscd}
\usepackage{bbm}
\usepackage{a4wide}

\theoremstyle{plain}

\theoremstyle{definition}

\newcommand{\ts}{\hspace{0.5pt}}

\begin{document}

\title[Scaling  and inverse scaling in anisotropic bootstrap percolation]
{Scaling and inverse scaling in anisotropic bootstrap percolation 
}
\author{Aernout C.D. van Enter}
\address{
Johann Bernoulli Institute for Mathematics and
  Computer Science, 
\newline
 \hspace*{\parindent}University of Groningen, 
  PO Box 407, 9700\ts AK Groningen, 
The Netherlands}
\email{a.c.d.van.enter@rug.nl}

\maketitle
{\bf Abstract:}

In bootstrap percolation it is known that the critical percolation threshold tends to converge slowly to zero with increasing system size, or, inversely, the critical size diverges fast when the percolation probability goes to zero. To obtain higher-order terms (that is, sharp and sharper thresholds) for the percolation threshold in general is a hard question. In the case of two-dimensional anisotropic models, sometimes such correction terms can be obtained from inversion in a relatively simple manner.  


\bigskip 


\bigskip

\section{Bootstrap percolation models}
 Bootstrap percolation models (also known in the literature as k-core percolation \cite{HS}, neuropercolation \cite{Koz,KPBBF}, jamming percolation \cite{Ton}, quorum percolation \cite{EMSTZ} or diffusion percolation \cite{AA})  are Cellular Automata, with a deterministic discrete-time  dynamics.  Often, however, probability is brought in, as one considers  probabilistic initial conditions. Although bootstrap percolation models are not PCA's in the proper sense, as CA's combined with probability, they are close relations of PCA's. 

Bootstrap percolation models describe the growth of  sets of occupied vertices (or sites) of a graph. At all vertices of a graph (whether finite or infinite) one places at an initial time  with probability $p$ a particle. The bootstrap percolation rule then requires each occupied vertex to stay occupied, and each empty vertex to become  occupied whenever sufficiently many vertices in its neighbourhood are occupied. The choice of graph, the choice of ``sufficiently many'' and the choice of the neighbourhood determine the model. One is interested whether after sufficiently many iterations each vertex gets occupied or not, and how this depends on the value of $p$. In particular one wants to know what happens for infinite graphs, or for sequences of increasing graphs. One also can consider more general rules, where an empty site gets occupied once a particular configuration, or one out of a particular set of configurations, in some neighbourhood  is occupied. E.g. the ``modified'' bootstrap percolation model requires that one neighbouring site along each lattice axis is occupied.  The bootstrap percolation models have some obvious monotonicity properties, in particular, the number of occupied vertices is growing in time, and there is stochastic monotonicity, in the sense that    the occupation number of each vertex of one evolving configuration is always larger or equal than that of  second evolving configuration once this holds at the beginning.

\smallskip

Bootstrap percolation models have been applied in a variety of contexts, e.g. for the study of metastability \cite{AL} and for  magnetic models \cite{Adl}, for the glass transition \cite{Ton} and for capillary fluid flow \cite{LZ}, for study of neural networks \cite{Amin,ET}, rigidity \cite{CRV}, contagion \cite{LV}, and they have also been  studied for purely mathematical interest, including recreational mathematics \cite{BolMem,NYT,Win1,Win2}.

Most interest is in the so-called critical models, in which the growth rule is such that  a finite set of occupied sites (= vertices) cannot fill the infinite lattice, and at the same time  all finite empty sets in an infinite occupied environment will be filled.

The simplest such models on (hyper-)cubic lattices are those where one considers the nearest neighbours of each site, and requires half of them to be occupied to get occupied at the next step, or where one requires at least one occupied site among the neighbours along every axis (modified bootstrap percolation).
In these models the most detailed results are known. In particular, it is known that on infinite lattices the percolation threshold is trivial ($p_c=0$), that is, for every positive $p$ every vertex of  the infinite lattice will in the end be occupied with probability one \cite{vE,Sch}. 

Moreover, on finite volumes the percolation threshold (now defined as the smallest value of $p$ above which the volume will be occupied with probability above one half) scales as a $d-1$-repeated logarithm of the size of the volume (that is, $p_c = O(\frac{1}{\ln V})$ in $d=2$, $p_c =O(\frac{1}{\ln \ln V})$ in $d=3$, etc). Such behaviour, with different constants for lower and upper bounds, was proven in  \cite{AL,CC,CM}, and with coinciding  lower and upper bounds in \cite{Hol1,Hol2, BBM, BBD-CM}. These last type of results (that is, 
$ p_c = C \frac{1}{\ln V} + o(\frac{1}{\ln V})$ in $d=2$, or 
$ p_c = C \frac{1}{\ln \ln V}+o(\frac{1}{\ln \ln V})$ 
in $d=3$, and similarly in general $d$ with $d-1$ times  repeated logarithms  for higher dimensions) 
have been called ``sharp thresholds''. 

As for lower-order corrections, (estimates on those are also called ``sharper'' thresholds in the literature),   in $d=2$ the $o(\frac{1}{\ln V})$ terms were shown to be of order $O( \frac{1}{(\ln V)^{\frac{3}{2}}})$, see \cite{Morr}. This strengthened earlier results of \cite{GH,GHM}. For higher-dimensional results about ``sharper''  thresholds, see \cite{Uzz}.
These sharper thresholds describe the systematic error which computational physicists in the past have run into, as is discussed e.g. in  \cite{AdLev,GLBD}

However, another notion of sharp thresholds, based on a sharp-threshold theorem of Friedgut and Kalai,  was presented in \cite{BB}. This $\varepsilon$-window 
-the window within which with large probability one will find the answer- provides an estimate for the statistical error, which is of order $O( \frac{\ln \ln V}{\ln^2 V})$ and hence much smaller than the systematic error. The statistical error being small with respect to the systematic error has been  a source for various erroneous numerical estimates of percolation thresholds and their numerical precision in the past, as errors tended to be substantially underestimated.  

In this contribution, I plan to describe to what extent the behaviour of bootstrap models is modified once the model becomes anisotropic, and in particular ``unbalanced'' (compare \cite{D-CH}). In particular, I will concentrate on the $(1,2)-$model, introduced in  \cite{GG1},  in which one considers an anisotropic  neighbourhood consisting of the  nearest neighbours along one axis, and the nearest and next-nearest neighbours along the other axis. 
 The distinction between balanced and unbalanced rules is that in balanced cases the growth occurs with the same speed in different directions, whereas in unbalanced cases there are easy and hard directions for growth. It appears to be the case that in $d=2$ a wide class of growth models is either balanced or unbalanced and that both classes display a characteristic scaling behaviour \cite{Morpriv}. 
 
 In higher dimensions it turns out that the leading behaviour is ruled by the two ``easiest'' growth directions \cite{EF}.

\section{ A tractable example: The (${1,2}$)-model}

In the (${1,2}$)-model the neighbourhood of each site in $Z^2$ consists of 2 sites in the East and West directions, and one site in the North and South directions.
In picture form:
\newline
\begin{tabular}{ccccccc}
\, & \, & \, & \, & $\bullet$ & \, & \, \\
$\mathcal{N}$ & $=$ & $\bullet$ & $\bullet$ & $0$ & 
$\bullet$ & $\bullet$ \\
\, & \, & \, & \, & $\bullet$ & \, & \,
\end{tabular}
\newline
At every step each empty site which has 3 of its neighbours (out of the 6 possible ones) occupied, becomes itself occupied, and every occupied site stays occupied forever. As an initial condition, we take a percolation configuration with initial occupation probability $p$.
This model, which is critical, was introduced by Gravner and Griffeath \cite{GG1} and they looked at its finite-size behaviour. The model is similar to, but somewhat easier to analyse than, the North-East-South model of Duarte \cite{Dua}, for which related but somewhat weaker results are known \cite{ADE,M}. 

The fact that $p_c=0$ in the infinite lattice follows from an argument due to Schonmann, first given for Duarte's model \cite{ADE,Sch2}. 

Indeed, let a $ 2 \times n$ rectangle be occupied, then the probability that this rectangle grows both Eastwards and Westwards is larger than the probability that at least 1 site in the columns East and West of this rectangle is occupied, which  is $[1 - (1-p)^n]^2$. 
The probability that this occurs in each column in a rectangle of size $l \times n$ we bound from below by $[1 - (1-p)^n]^l$. Choose $n=  \frac{C}{p} \ln \frac{1}{p}$, then this probability can be bounded by $(1-p^C)^l$; once $C \geq 2$ and $l \geq \frac{1}{p^C}$,  such a rectangle keeps growing in both directions with large probability; the fact that such an occupied and growing rectangle can occur with positive probability implies that somewhere in an infinite lattice such a nucleation center will occur, and it will then fill up the whole lattice. 

The question after this is how big a square volume should be for such a nucleation center to occur with large  probability (e.g. probability a half).
The argument given above predicts that a $2 \times n$ rectangle occurs at some fixed location with probability at least  $p^{2n} = e^{-O(\frac{1}{p} \ln^2 \frac{1}{p})}$, 
and that therefore the size of the square volume $V= N \times N$ should be the inverse of that probability, that is,  if $N$ (or $V) \geq  e^{+O(\frac{1}{p} \ln^2 \frac{1}{p})}$, it can be filled with large probability. Inverting the argument implies an upper bound for the rate at which  the percolation threshold decreases as a function  of $V$, of the form 
\begin{equation}
p_c \leq C_1 \frac{\ln^2 \ln V}{\ln V}.
\end{equation}    

An argument providing a lower bound for $p_c$ of the same order, that is 
\begin{equation}
p_c \geq C_2 \frac{\ln^2 \ln V}{\ln V}.
\end{equation}    
was developed in \cite{EH}, using and correcting the analysis of \cite{GG1}.

In fact, one can improve the on above strategy, as follows. (See \cite{DEHM}, following \cite{DE}).\\ 
One starts with a $2 \times  \frac{2}{p} \ln \ln \frac{1}{p}$  rectangle, which has all its even (or odd) sites occupied, then at the next step, the whole rectangle is filled. After that, one grows with vertical steps of size $1$ and horizontal steps of increasing size, through a sequence of rectangles $R_n$, which in the $y$-direction have size $n$ and in the $x$-direction have size $ \frac{1}{3p} \exp 3np $. This goes on until we reach the size $n= \frac{1}{3p} \ln \frac{1}{p}$. With this choice the probability for a rectangle $R_n$ to  grow a step in the $x$-direction equals the probability to  grow a step in the $y$-direction. 

The  probability of growing a step in the vertical direction from a rectangle $R_n$ is approximately $8p^2 x_n$ (one needs two occupied sites close enough, the factor 8 here is of combinatorial origin) which equals $\frac{8p}{3} \exp 3np$.
The probability of growing in the horizontal direction, over a distance 
$x_{n+1} -x_n$ equals a constant term $\frac{1}{e}$, for every $n$.

One thus needs to compute the product from $n_0= \frac{2}{p} \ln \ln \frac{1}{p}$ to $n_f= \frac{1}{3p} \ln \frac{1}{p}$ over these probabilities.\\
The result is

\begin{equation}
\begin{split}
\prod_{n=n_0}^{n=n_f} \frac{8p}{3e} \exp 3np 
& = {\frac{8p}{3e}}^{(n_f- n_0)} \exp[ 3p \sum\limits_{n=n_0}^{n=n_f} n] = {\frac{8p}{3e}}^{(n_f- n_0)} \exp[ 3p (\frac{1}{2} n_f(n_f-1) - \frac{1}{2}n_0(n_0 -1))] \\
& = \exp [-\frac{1}{6p} \ln^2 \frac{1}{p} + \frac{1}{3} \ln \frac{8}{3e} \frac{1}{p}\ln \frac{1}{p} + o(\frac{1}{p}\ln \frac{1}{p})].     
\end{split}
 \end{equation}

This is our main result, for the detailed proof that this strategy indeed provides the best estimate, see \cite{DEHM}.

\section{Inversion}
If the probability for a nucleation center to occur at a fixed location is given by an expression of the form 
$P= \exp -\frac{C}{p}$, the necessary volume size to see such a nucleation center with substantial probability in that volume, that is the ``critical volume''  will be $V_c = \exp + \frac{C}{p}$, which is easily inverted, resulting  in an expression of the form  $p_c =  C \frac{1}{\ln V}$ for the critical percolation threshold as a function of the volume.

However, if there are logarithmic corrections and subdominant terms as above, that is 
\begin{equation}
V_c  = \exp [ \frac{C}{p} \ln^2 \frac{1}{p} +  \frac{C'}{p} \ln \frac{1}{p}],
\end{equation}
to invert such expressions we need to perform some extra steps. We observe the following:

\begin{equation}
p_c =  \frac{1}{\ln V} (C \ln^2\frac{1}{p_c} +C' \ln \frac{1}{p_c}).
 \end{equation}

We also notice that in the limit of $V$ large and hence $p_c$ small it holds that 
\begin{equation} 
\frac{1}{p_c} \leq \ln V \leq \frac{1}{p_c^{1+ \varepsilon}}
\end{equation}
and (by taking logarithms)
\begin{equation} 
\ln \frac{1}{p_c} \leq \ln \ln V \leq (1 + \varepsilon)\ln \frac{1}{p_c}
\end{equation}
and 
\begin{equation} 
\ln \ln \frac{1}{p_c} \leq \ln \ln \ln V \leq \ln \ln \frac{1}{p_c} + \varepsilon .
\end{equation}
Thus asymptotically, by substitution plus using the above estimates 
\begin{equation} 
\begin{split}
p_c & =  \frac{1}{\ln V} [C \ln^2\frac{1}{p_c} +C' \ln \frac{1}{p_c}]\\
& = \frac{1}{\ln V} [C (\ln \ln V  - 2 \ln \ln \ln V - \ln C +O(\varepsilon))^2 +C' \ln \ln V + O(\varepsilon)]\\
& = \frac{1}{\ln V} [C (\ln^2 \ln V  - 4 \ln \ln \ln V \ln \ln V- 2\ln C \ln \ln V) + +C' \ln \ln V + O(\ln^2 \ln \ln V)].
\end{split}
\end{equation}
Hence knowing the second term in the critical volume provides a third term in the critical probability, and we also notice that the second term in the critical probability does not depend on the constant $C'$ of this second critical-volume term.

A related argument was used in \cite{Susan} to estimate the $\varepsilon$-window. This analysis extended the analysis of \cite{BB}, applying the sharp-threshold theorem of Friedgut and Kalai, and  the $\varepsilon$-window turns out to have width $O(\frac{\ln^3 \ln V}{\ln^2 V})$.

Numerically, that is for computational physicists e.g.,  these results are totally discouraging. Whereas in standard bootstrap percolation to obtain a $99 \%$ accuracy in $p_c$   the order of magnitude of a square already needs to be of order $O(10^{3000})$ \cite{GH}, in the $(1,2)$-model one needs to go even higher, namely to a doubly exponential size of order $O(10^{10^{1400}})$. These findings illustrate the point made in \cite{Gray}, that Cellular Automata, despite being discrete in state, space, and time,  may still be ill-suited for computer simulations.  

\section{Generalisations: related models, higher dimensions and other graphs}
In ordinary and modified bootstrap percolation we have quite precise results.
There is a variety of related models with similar behaviour, e.g. \cite {BM1, BM2, Frob,HLR}. In particular it is remarkable that the model of \cite{BM2} is anisotropic, but nonetheless scales in the same way as ordinary bootstrap percolation; in the terms of \cite{D-CH} it is ``balanced''. A much wider class of models was  recently considered in \cite{BSU}, in which some general order-of-magnitude results were derived for critical models. More recently \cite{Morpriv} it was shown that this class consists of two subclasses, either the balanced ones, such as ordinary bootstrap percolation, which display similar asymptotic behaviour, or the unbalanced ones, in which logarithmic corrections of the type displayed in the $(1,2)$-model occur. The essential distinction is that balanced models grow at the same rate in two different directions, whereas unbalanced models have an easy and a hard growth direction.

There exist also some results on bootstrap percolation  in higher dimensions. In the anisotropic case, for the time being we only know order-of-magnitude results for $(a,b,c)$-models, in which neighbourhoods are considered which consist of neighbours at distances $a,b$ and $c$ ($a \leq b\leq c$) along the different axes \cite{EF} (of which again half the sites need to be occupied to occupy an empty site). The result is that the scaling becomes doubly exponential c.q. doubly logarithmic, similarly to the isotropic models \cite{CC,CM}, but with the behaviour controlled by the two-dimensional $(a,b)$-model. One bound is based on a variation of Schonmann's \cite{Sch} induction-on-dimension argument, the other direction follows a similar strategy as \cite{CC}. To establish any form of a sharp threshold, however,  is open for the time being.\\
In two dimensions the $(1,b)$-models can be analysed along  similar lines as the $(1,2)$-model, which results in the same asymptotics, but  with the (sharp and computable) constant \\
$C=\frac{(b-1)^2}{2(b+1)}$, rather than $C=\frac{1}{6}$, as the leading term. To establish such a result for the Duarte model, however, remains open. 

A quite different family of results, in which there is a transition at a finite threshold $p$, occurs for bootstrap percolation models on either trees \cite{BBP,BGHJP,CLR,SLC}, random graphs \cite{BP,PSW}, or hyperbolic lattices \cite{STBT}.
Such transitions have a ``hybrid'' (mixed first-second order) character, in the sense that on the one hand, one finds that at the threshold the infinite cluster has a minimum density (so it jumps from zero, just as one expects at a first-order phase transition), while at the same time there are divergent correlation lengths and  non-trivial critical exponents, which are characteristic for second-order (critical) phase transitions. Such hybrid ``random first order'' transitions have been proposed to be characteristic for glass transitions. See e.g. \cite{het}. On regular lattices models with this kind of behaviour cannot be constructed via the type of bootstrap percolation rules discussed above, but more complicated Cellular Automaton growth rules  with this type  of behaviour have been studied in \cite{JS,TB}.

\smallskip

{\bf Acknowledgements:} I thank the organisers for their invitation to talk at the EURANDOM meeting on Probabilistic Cellular Automata, and I thank my colleagues and coworkers, Joan Adler, Jose Duarte, Hugo Duminil-Copin, Anne Fey-den Boer, Tim Hulshof and Rob Morris, as well as Susan Boerma-Klooster and Roberto Schonmann,  for all they taught me. I thank Rob Morris for correcting me on the $(1,b)$-constant. Moreover I thank Tim Hulshof for helpful advice on the manuscript.


\begin{thebibliography}{99}
\small

\bibitem{AA} J.~Adler and A.~Aharony, Diffusion Percolation: infinite time limit and bootstrap percolation. {\bf J. Phys. A 21}, 1387--1404 (1988).

\bibitem{Adl}J.~Adler, Bootstrap percolation. {\bf Physica A 171}, 452--470 (1991).

\bibitem{AdLev}J.~Adler and U.~Lev, bootstrap percolation: visualisations and applications. {\bf Braz. J. Phys. 33}, 641--644 (2003).

\bibitem{ADE} 
J. Adler, J.A.M.S. Duarte and A.C.D. van Enter,
Finite-size effects for some bootstrap percolation models.
{\bf J. Stat. Phys.} \textbf{60}, 323-332 (1990), and addendum
{\bf J. Stat. Phys.} \textbf{62}, 505--506 (1991).

\bibitem{AL} M.~Aizenman and J.~L.~Lebowitz, Metastability effects in bootstrap percolation. {\bf J. Phys. A 21}, 3801--3813 (1988).

\bibitem{Amin} H.~Amini, Bootstrap percolation in living neural networks. {\bf J. Stat. Phys. 141}, 459--475 (2010). 


\bibitem{Susan} Susan Boerma-Klooster, A sharp threshold for an anisotropic bootstrap percolation model. Groningen bachelor thesis (2011). 

\bibitem{BB} J.~Balogh and B.~Bollobas, Sharp thresholds in bootstrap percolation. {\bf Physica A 326}, 305--312 (2003). 

\bibitem{BBM}
J. Balogh, B. Bollobas and R. Morris:
 Bootstrap percolation in three dimensions.
{\bf Ann. Prob.} \textbf{37}, 1329--1380 (2009).

\bibitem{BBD-CM}
J. Balogh, B. Bollobas, H. Duminil-Copin and R. Morris,
  The sharp threshold for bootstrap percolation in all dimensions.
arXiv:1010.3326v1 (2010), {\bf Trans. Am. Math. Soc.} \textbf{364}, 2667--2701 (2012). 

\bibitem{BBP} J. Balogh, B. Bollobas and G. Pete, Bootstrap percolation on infinite trees and non-amenable groups. {\bf Combin. Prob. Computing} \textbf{15}, 715--730 (2006)

\bibitem{BP} J. Balogh and B.~G. Pittel, Bootstrap Percolation on the  Random Regular Graph. {\bf Random Struct. Alg.} \textbf{30}, 257--286 (2007). 

\bibitem{BGHJP} B. Bollobas, K. Gunderson, C. Holmgren, S. Janson, M. Przykucki,
Bootstrap percolation on Galton-Watson trees.
{\bf El. J. Prob.} \textbf{19}, 13 (2014).


\bibitem{BolMem} B. Bollobas, The Art of mathematics: Coffee Time in Memphis, problems 34 and 35, Cambridge University Press, Cambridge (2006).

\bibitem{BSU} B. Bollobas, P. Smith, A. Uzzell, Neighborhood family percolation
(original title: Generalized bootstrap percolation). arXiv:1204.3980v2 (2012). 

\bibitem{BM1} K.~Bringmann and K.~Mahlburg, Improved bounds on metastability thresholds and probabilities for generalized bootstrap percolation. {\bf Trans. Am. Math. Soc.} \textbf{364}, 3829--2859 (2012).

\bibitem{BM2}K.~Bringmann,  K.~Mahlburg and A.~Mellit, Convolution bootstrap percolation models, Markov-type stochastic processes, and mock theta functions.
{\bf Int. Math. Res. Notices 2013}, 971--1013 (2013).

\bibitem{CC}
R. Cerf and E. Cirillo,
 Finite size scaling in three-dimensional bootstrap percolation.
{\bf Ann. Prob}. \textbf{27}(4), 1837--1850 (1999).

\bibitem{CLR}
J. Chalupa, P.L. Leath and G.R. Reich, Bootstrap percolation on a Bethe lattice. {\bf J. Phys. C} \textbf{12}, L31--L35 (1979).


\bibitem{CM}
R. Cerf and F. Manzo,
 The threshold regime of finite volume bootstrap percolation. 
{\bf Stoch. Proc. Appl.} \textbf{101}, 69--82 (2002).

\bibitem{CRV}R.~Connelly, K.~Rybnikov and S.~Volkov, Percolation of the loss of tension in an infinite triangular lattice. {\bf J. Stat. Phys. 105}, 143--171 (2001).

\bibitem{Dua} 
J.A.M.S. Duarte, Simulation of a cellular automaton with an oriented bootstrap rule. \newblock {\bf Physica A} \textbf{157}, 1075--1079 (1989).

\bibitem{DE}
H.~Duminil-Copin, A.C.D. van Enter,
Sharp metastability threshold for an anisotropic bootstrap percolation model. {\bf Ann. Prob. 41}, 1218--1242 (2013).

\bibitem{DEHM}
H.~Duminil-Copin, A.C.D. van Enter, W.J.T. Hulshof and R.~Morris, in preparation.

 \bibitem{D-CH}H.~Duminil-Copin and A.E.~Holroyd, Finite volume bootstrap percolation with balanced threshold rules on $Z^2$. Preprint, obtainable at 
http:{//}www.unige.ch{/}duminil{/} (2012).


\bibitem{EMSTZ} J.-P.~ Eckman, E.~Moses, E.~Stetter, T.~Tlusty and C.~Zbinden, Leaders of neural cultures in a quorum percolation model. {\bf Front. Comput. Neurosci.
4}, 132 (2010).

\bibitem{ET} J.-P. Eckman and T. Tlusty. Remarks on bootstrap percolation in metric networks. {\bf J.Phys. A: Math. Theor. 42 }, 205004 (2009). 

\bibitem{vE}
A.C.D. van Enter,
Proof of Straley's Argument for Bootstrap 
Percolation.
{\bf J. Stat. Phys.} \textbf{48},  943--945 (1987).

\bibitem{EF}
A.C.D.~ van Enter and A.~Fey,
Metastability Thresholds for Anisotropic Bootstrap Percolation in Three Dimensions.
{\bf J. Stat. Phys.} \textbf{147},  97--112 (2012).

\bibitem{EH}
A.C.D. van Enter and W.J.T. Hulshof,
 Finite-size effects for anisotropic bootstrap percolation: logarithmic corrections. 
{\bf J. Stat. Phys.} \textbf{128}, 1383--1389 (2007).

\bibitem{Frob} K.~Frob\"ose, Finite-size Effects in a Cellular Automaton for Diffusion. 
{\bf J. Stat. Phys.} \textbf{55}, 1285--1292 (1989).

\bibitem{GG1}J.~Gravner and D.~Griffeath, First passage times for threshold growth dynamics on $Z^2$. {\bf Ann. Prob. 24}, 1752--1778 (1996).

\bibitem{GH} J.~Gravner and A.E.~Holroyd, Slow convergence in bootstrap percolation. {\bf Ann. Appl. Prob. 18}, 909--928 (2008).

\bibitem{GHM} J.~Gravner, A.E.~Holroyd and R.~Morris, A sharper threshold for bootstrap percolation in two dimensions. {\bf Prob. Th. Rel. Fields 153}, 1--23 (2012).

\bibitem{Gray}
L.~Gray, A Mathematician looks at Wolfram's new kind of science.
{\bf Notices of the Am. Math. Soc. 50}, 200--211 (2003).
 
\bibitem{GLBD} P.~de~Gregorio, A.~Lawlor, P.~Bradley, K.A.~Dawson, Clarification of the bootstrap percolation paradox. {\bf Phys. Rev. Lett. 93}, 025501 (2004). 

\bibitem{HS} A.B.~ Harris and J.M.~Schwarz, $ \frac{1}{d}$ expansion for $k$-core percolation. {\bf Phys. Rev. E 72}, 046123 (2005).

\bibitem{het} Dynamic Heterogeneities in glasses, Colloids, and Granular Media.
Eds. L.~Berthier, J.P.~Bouchaud, L.~Cipelletti and W.~van Saarloos. Oxford University Press, Oxford-New York (2011).

\bibitem{Hol1} A.E.~Holroyd, Sharp metastability threshold for two-dimensional bootstrap percolation. {\bf Prob. Th. Rel. Fields 125}, 195--224 (2003).

\bibitem{Hol2} A.E.~Holroyd,
The metastability threshold for modified bootstrap percolation in d dimensions.
{\bf Electron. J. Probab.} \textbf{11}, no. 17, 418--433 (2006).

\bibitem{HLR} A.E.~Holroyd, T.M.~Liggett and D.~Romik, Integrals, partitions and cellular automata. {\bf Trans. Am. Math. Soc.} \textbf{356}, 3349--3368 (2004).

\bibitem{JS} M.~Jeng and J.~M.~Schwarz, On the study of jamming percolation. {\bf J. Stat. Phys.} \textbf{131}, 575--595 (2008).

\bibitem{Koz} R.~Kozma, Neuropercolation. 
http:{//}www.scholarpedia.org{/}article{/}Neuropercolation

\bibitem{KPBBF} R.~Kozma, M.~Puljic, P.~Balister, B.~Bollobas and W.J.~Freeman, Phase transitions in the neuropercolation model of neural populations with mixed local and non-local interactions. {\bf Biol. Cybern. 92}, 367--379 (2005).

\bibitem{LV} I.H.~Lee and A.~Valentiniy, Noisy contagion without mutation. {\bf Rev. Econ. Studies 67}, 47--67 (2000).

\bibitem{LZ} R. Lenormand and C. Zarcone, Growth of clusters during imbibition in a network of capillaries. {\bf Kinetics of Aggregation and Gelation}, Eds. F. Family and D.P. Landau, Elsevier, Amsterdam p177--180 (1984).


\bibitem{Morr} R.~ Morris, The second term for bootstrap percolation in two dimensions. Manuscript in preparation. See 
http:${/}{/}$w3.impa.br${/}$~rob${/}$index.html

\bibitem{Morpriv} R.~Morris (private communication) has informed me that this has been proven in a joint work, in preparation,  with B.~Bollobas, H.~Duminil-Copin and P.~Smith (2014).


\bibitem{M} T.S.~Mountford, Critical lengths for semi-oriented bootstrap percolation. {\bf Stoch. Proc. Appl. 95}, 185--205 (1995).



\bibitem{NYT} New York Times Wordplay Blog, 8 July 2013, see\\
http:${/}{/}$wordplay.blogs.nytimes.com${/}2013{/}07{/}08{/}$bollobas${/}?${\_}$r=0$ 


\bibitem{PSW} B. Pittel, J. Spencer and N. Wormald, Sudden Emergence of a Giant k-Core. {\bf J. Comb. Th. B} \textbf{67}, 111--151 (1996). 

\bibitem{STBT} F. Sausset, C. Toninelli, G. Biroli an G. Tarjus, Bootstrap Percolation and Kinetically Constrained Models on Hyperbolic Lattices.
{\bf J. Stat. Phys.} \textbf{138}, 411--430 (2010).  


\bibitem{Sch} R.H.~Schonmann, 
On the Behavior of some Cellular Automata related to Bootstrap Percolation.
{\bf Ann. Prob.} \textbf{20}, 174--193 (1992).

\bibitem{Sch2} R.H. Schonmann,
Critical points of 2-dimensional bootstrap 
percolation-like cellular automata.
{\bf J. Stat. Phys.} {\bf 58}, 1239--1244 (1990). 

\bibitem{SLC}
J.M. Schwarz, A.J.~Liu and L.Q. Chayes, The onset of jamming as the sudden emergence of an infinite k-core cluster. {\bf Europhysics Lett.} \textbf{73}, 560--566 (2006). 

\bibitem{TB} C.~Toninelli and G.~Biroli, A new class of cellular automata with a discontinuous glass transition. {\bf J. Stat. Phys.} \textbf{130}, 83--112 (2008).

\bibitem{Ton} C.~Toninelli, Bootstrap and jamming percolation. In: J.~P.~Bouchaud, M.~M\'ezard, J.~Dalibard (eds.) Les Houches school on Complex Systems, Session LXXXV, 289--308 (2006).

\bibitem{Uzz} A.E.~Uzzell, An improved upper bound for bootstrap percolation in all dimensions. arXiv 1204.3190 (2012). 

\bibitem{Win1}
P. Winkler, Mathematical Puzzles (p 79, The infected Checkerboard).
A.~K. Peters Ltd (2004).

\bibitem{Win2}
P. Winkler, Mathematical Mindbenders (p 91, Infected Hypercubes), A.~K. Peters Ltd (2007).

 

\end{thebibliography}
\end{document}